\documentstyle[aps,psfig,eqsecnum]{revtex}
\begin{document}
\twocolumn

\def\vm{v_{\rm max}}
\def\beq{\begin{equation}}
\def\eeq{\end{equation}}
\def\la{\langle}
\def\ra{\rangle}
\def\k{\kappa}
\def\a{\alpha}
\def\b{\beta}
\def\d{\delta}
\def\o{\omega}
\def\gsim{\; \raisebox{-.8ex}{$\stackrel{\textstyle >}{\sim}$}\;}
\def\lsim{\; \raisebox{-.8ex}{$\stackrel{\textstyle <}{\sim}$}\;}
\def\gtrsim{\gsim}
\def\lessim{\lsim}
\def\g*{g_{*\mu\nu}}
\def\gjf{\widetilde{g}_{\mu\nu}}
\def\Gt{\widetilde{G}}
\def\Mt{\widetilde{M}}
\def\Lt{\widetilde{L}}
\def\phi{\varphi}

\title{Primordial black hole evolution in tensor-scalar cosmology}

\author{ 
Ted Jacobson\thanks{E-mail: jacobson@physics.umd.edu} }
\address{Department of Physics, University of Maryland,
College Park, MD 20742-4111\\
and\\
Institute for Theoretical Physics, University of California, 
Santa Barbara, CA 93106}
\maketitle
\begin{abstract}
A perturbative analysis shows that black holes do not remember the 
value of the scalar field $\phi$ at the time they formed if $\phi$ changes
in tensor-scalar cosmology. Moreover, even when the black hole mass 
in the Einstein frame is approximately unaffected by the changing of $\phi$,
in the Jordan-Fierz frame the mass increases. This mass increase requires 
a reanalysis of the evaporation of primordial black holes in tensor-scalar 
cosmology. It also implies that there could have been a significant 
magnification of the (Jordan-Fierz frame) mass of primordial black holes.
\end{abstract}
  
\section{INTRODUCTION}

Various approaches to unified theories and quantum gravity
suggest the possibility of one or more massless 
scalar fields such as the dilaton or other moduli coupled to
the trace of the energy-momentum tensor of matter. 
Theories with such fields in addition to the spacetime
metric are dubbed ``tensor-scalar" theories. 
The action for an illustrative  
class of such theories with one scalar 
is\cite{DamourEsposito}
\begin{eqnarray}
S&=&(16\pi G_*)^{-1}\int d^4x\, g_*^{1/2}
\Bigl(R_*-2g_*^{\mu\nu}\partial_\mu\phi\partial_\nu\phi\Bigr)\nonumber\\
&& +S_m[\psi_m,A^2(\phi)g_{*\mu\nu}].
\label{S}
\end{eqnarray}
The metric $g_{*\mu\nu}$ is called the ``Einstein metric" and $R_*$ is
its scalar curvature. The matter fields are collectively denoted
$\psi_m$, and they couple universally to the ``Jordan-Fierz metric"
\beq
\gjf\equiv A^2(\phi)\g*,
\label{gjf}
\eeq
where the form of the coupling function $A(\phi)$ is input 
which presumably descends from a more fundamental theory.
The Newton constant $\widetilde{G}$ 
as measured in Cavendish type experiments is given by 
\beq
\widetilde{G}=(1 + \a^2(\phi)) A^2(\phi)G_*,
\label{Gt}
\eeq
where $\a\equiv d\ln A/d\phi$,
which is not in fact constant in a cosmological solution.
Quantities defined with respect to
$g_{*\mu\nu}$ are commonly referred to as being given 
in the ``Einstein frame" while those defined with respect to
$\gjf$ are said to be in the ``Jordan-Fierz frame".

Damour and Nordtvedt\cite{DamourNordtvedt} identified a generic attractor 
mechanism which drives a 
tensor-scalar cosmology to a purely tensor (Einstein) one at
late times, thus significantly increasing the plausibility of 
these models. According to this mechanism the scalar $\phi$ 
is driven to a local minimum of $A(\phi)$.  
Definite predictions for the residual effects of 
the scalar(s) emerge from their analysis.
The Jordan-Fierz-Brans-Dicke theory is a special case of (\ref{S})
with $A(\phi)=\exp(\alpha_0\phi)$. Since such a coupling function has no
minimum the attractor mechanism does not work for this theory and
fine tuning is therefore required. Generically, however, one might
expect $A(\phi)$ to possess local minima. A related mechanism in
a model motivated by string theory was studied by Damour and 
Polyakov\cite{DamourPolyakov}.

Barrow and Carr initiated a study of the evolution
of a population of primordial black holes in tensor-scalar
cosmology\cite{Barrow,BarrowCarr}.
They considered two differing scenarios\cite{Barrow}, ``scenario A" in 
which the value of the scalar field (and hence $\widetilde{G}$)
at the horizon evolves along with the cosmological value,
and ``scenario B" in which
a black hole ``remembers" the value of the scalar field at the
time of its formation. The 
black hole evolution is very different in the two cases if, as is
natural in these models,
Newton's constant (\ref{Gt}) decreases over cosmological timescales.
Analyzing the black hole evolution in the Jordan-Fierz frame,
and assuming the black hole mass $\Mt$ changes only due to the
Hawking evaporation, Barrow and Carr argued that
in scenario A the Hawking luminosity
increases as $L\sim T_H^4 \times({\rm Area})\sim (\Gt \Mt)^{-2}$. 
A black hole born when
$\Gt$ was larger would thus have a longer lifetime than would be 
surmised from the present value of $\Gt$. In scenario B 
on the other hand,
the black hole remembers the primordial value of $\Gt$ so
its lifetime would be even longer. 

It turns out that neither of these two scenarios is correct.
It will be shown in this paper 
first that there is no ``gravitational memory", so
the value of $\Gt$ at the black hole keeps up with the cosmological change.
Second, even if the mass of the black hole is essentially constant 
in the Einstein
frame, the mass {\it increases} 
in the Jordan-Fierz frame in proportion
to $1/A(\phi)$. This mass increase would
counteract the Hawking evaporation 
(as described in the Jordan-Fierz frame),
and could significantly ``magnify"
the mass of primordial black holes.

\section{Evolution of the scalar field at the horizon}

The problem to solve is this: if a small black hole is embedded in 
a cosmology with changing scalar field 
how does the scalar field at the horizon evolve? At the outset,
it is worth remarking that it seems unlikely that the scalar 
field would be pinned at the horizon, since that would entail increasing
gradients in the scalar field as it interpolates between the horizon and
the changing cosmological value. It would seem that such gradients
would lead to propagation that would even out the field. 

We approach the problem by exploiting the 
great separation of scales between the black hole and the cosmological
background. Since  the black hole is much smaller than the cosmological 
length or time scales it is reasonable to think of it as
sitting in a local asymptotically flat space, with a boundary condition
for the scalar field set by the cosmological evolution $\phi_c(t)$. If 
the scalar field at the black hole follows $\phi_c(t)$ 
then, since this 
change is very slow compared with the size of the black hole, the solution
should be a small perturbation of the stationary black hole. 
If such a perturbation exists 
in which $\phi$ at the horizon keeps up with $\phi_c(t)$, 
then our assumption will be shown to be consistent. 

For simplicity we first discuss nonrotating 
black holes,
and we work in the Einstein frame.
The only such black holes
in Einstein-scalar gravity are the Schwarzschild
metric with a constant scalar, 
and the only spherically symmetric perturbations of these
black holes are pure scalar fields satisfying the wave equation
in the Schwarzschild background. Thus we need only look for 
spherically symmetric solutions to the wave equation,
\beq
\Bigl(-g^{tt}\partial_t^2 + \frac{1}{\sqrt{-g}}\partial_r
\sqrt{-g}g^{rr}\partial_r\Bigr)
\phi(t,r)=0,
\label{wave}
\eeq
(in Schwarzschild coordinates)
subject to the boundary condition 
\beq
\phi(t,r=\infty)=\phi_c(t)=\dot{\phi_c}t.
\label{bc}
\eeq
Since the cosmological timescale is assumed to be very long compared
with that of the black hole, it is consistent to set the 
cosmological scalar perturbation equal to a linear function
of the Schwarzschild time coordinate $t$ as we have done here
with $\dot{\phi}_c$ a constant.

The unique solution to the wave equation (\ref{wave}) depending only
on $t$ and satisfying the boundary condition is  
just the boundary value itself,
\beq
\phi_1(t,r)=\dot{\phi}_c t.
\label{phi1}
\eeq
However this can not be the perturbation we are looking for 
since the coordinate $t$  and hence the ``perturbation"
$\phi_1$ diverges at the black hole horizon.  The unique
solution depending only on $r$ is given up to constants by
\beq
\phi_2(t,r)= \ln(1-r_0/r)
\label{phi2}
\eeq
where $r_0$ is the Schwarzschild radius and we have used  
$\sqrt{-g}g^{rr}=(r-r_0)r\sin\theta$.
This solution also diverges at the horizon, but there is
a linear combination of these two solutions that is regular
at the horizon.\footnote{This observation is due to Amos Ori.}
Since $\phi_2$ vanishes at infinity, this linear combination
will be the perturbation we seek.

The advanced time coordinate $v=t+r^*$, with 
$r^*=r+r_0\ln(r/r_0-1)$,
is regular on the horizon.
In terms of $v$ we have  
$\ln(1-r_0/r)=(v-t-r)/r_0 + \ln r_0/r$, 
hence the linear combination of $\phi_1$ and $\phi_2$
which is regular on the horizon and approaches $\phi_c$ at infinity is 
\beq
\phi_3=\phi_1+r_0\dot{\phi}_c\phi_2 = \dot{\phi}_c(v-r-r_0\ln \frac{r}{r_0}). 
\label{phi3}
\eeq
 
The existence of the solution (\ref{phi3}) 
establishes the result that the
cosmological change of the scalar field can extend smoothly
to the black hole horizon. Any other regular perturbation satisfying 
the boundary condition must be a superposition of waves which
will dissipate by falling into the black hole or spreading
out to infinity, hence after transients the perturbation
(\ref{phi3}) will describe the secular change of the scalar field.

A surface of constant $\phi_3$ 
intersects the horizon at an advanced time $v_H$
and reaches infinity at a
cosmological or Schwarzschild time $t_\infty$,
these two times being related by $v_H-r_0=t_\infty$.
The two surfaces $v=v_H$ and $t=t_\infty$ thus intersect
at $r^*=v_H-t_\infty=r_0$ which is at around $r\simeq1.5r_0$.
Thus there is not much ``lag"
between the horizon value and the cosmological value.

The solution $\phi_3$ generalizes with little change to the 
case of a rotating black hole. Using Boyer-Lindquist coordinates
for the Kerr metric the wave equation for functions of $t$ and $r$
again takes the form (\ref{wave}). Hence
we again find that $\phi_1(t)$ (\ref{phi1})
solves the wave equation, and in place of (\ref{phi2}) we find the
stationary solution 
\beq
\phi_{2'}=\ln\frac{r-r_+}{r-r_-}
\label{phi2'}
\eeq
where $r_\pm$ are the radii of the outer and inner horizons
and we have used  $\sqrt{-g}g^{rr}=(r-r_+)(r-r_-)\sin\theta$.
The advanced time coordinate $v=t+r^*$,
now with 
$r^*=\int dr (r^2+a^2)/(r-r_+)(r-r_-) =  r+ \frac{r_++r_-}{r_+-r_-}
[r_+\ln(\frac{r}{r_+}-1)-r_-\ln(\frac{r}{r_-}-1)]$,
is regular on the Kerr horizon.
The linear combination of $\phi_1$ and $\phi_{2'}$
which is finite on the horizon and approaches
$\phi_c$ at infinity is given by
\begin{eqnarray}
\phi_{3'}&=&\phi_1+
\left(\frac{2mr_+}{r_+-r_-}\dot{\phi}_c\right)\phi_{2'} \nonumber\\
&=& \dot{\phi}_c\left[v-r-2m \ln(\frac{r}{r_-}-1)+
\frac{2mr_+}{r_+-r_-} \ln\frac{r_+}{r_-}\right]
\label{phi3'}
\end{eqnarray}
where $2m=r_++r_-$. Thus also for a rotating black hole there
is a perturbation describing a scalar field changing linearly
with time in the same way at the horizon as at infinity.

\section{Evolution of the black hole mass}

The mass of a black hole may grow by accretion or decrease
by emission of Hawking radiation. Aside from standard accretion
processes, in tensor-scalar cosmology a 
changing scalar at the horizon entails a flux of energy
into the black hole, thus increasing its mass. In addition,
we shall see that even if the mass in the Einstein frame is
approximately constant, the mass defined in the Jordan-Fierz frame changes
simply due to the conformal scaling of the  metric (\ref{gjf}).

To avoid confusion, in this section asterisk subscripts are attached
to quantities defined with respect to the Einstein metric,
and tildes adorn quantities defined with respect to the 
Jordan-Fierz metric. This asterisk is unrelated to the 
superscript on the
radial tortoise coordinate in the previous section.

The rate of change of a black hole mass with respect to 
cosmological time, arising from the changing scalar, 
is most easily evaluated in the Einstein frame. 
Let us consider for simplicity a nonrotating black hole.
Provided the change of the mass is slow on the scale of the
black hole, i.e. $dM_*/dt_*\ll M_*/r_{*0}$,
it is given approximately
by the flux of Killing energy across the horizon,
\beq
dM_*/dt_* ={\rm Area}_* \times T^{(\phi)}_{*~v_*v_*}=
2G_*^{-1} r_{*0}^2 (d\phi_c/dt_*)^2.
\eeq
If $\phi$ changes by an amount $\Delta\phi$ over an
Einstein-frame cosmological time $\Delta t_*$, the fractional
increase of the mass is given by 
\beq
\frac{\Delta M_*}{M_*}\sim \frac{r_{*0}}{\Delta t_*}(\Delta\phi)^2
\eeq 
provided $r_{*0}\Delta\phi/\Delta t_*\ll 1$. Small black holes
can therefore experience a huge variation in $\phi$ with essentially
no increase of the mass $M_*$ in the Einstein frame as long as the
change is adiabatic. This conclusion also follows directly
from the fact that in adiabatic processes the black hole entropy 
$A_*/4G_*=4\pi G_*M_*^2$ is unchanged.

The relation between the mass $\widetilde{M}$ in the Jordan-Fierz frame 
and that in the Einstein frame is given by 
$\widetilde{M}/M_*=dt_*/d\widetilde{t}$
(since the mass (energy) 
is by definition the value of the generator of (asymptotic) 
Killing time translations, which scales inversely with the 
time coordinate), which from (\ref{Gt}) yields 
\beq
\Mt=A^{-1}(\phi)M_*.
\label{Mt}
\eeq 
Thus, even when $M_*$ is constant, 
the black hole mass in the Jordan-Fierz frame is 
{\it not constant}. 
This conclusion is somewhat surprising since, by contrast, 
the rest mass of ordinary matter {\it is} constant in the Jordan-Fierz
frame.\footnote{In the string motivated variation\cite{DamourPolyakov}
 of (\ref{S}), however,  one expects
dilaton-dependence of (at least) hadron masses in both the Jordan-Fierz 
and Einstein frames.}

In the Einstein frame 
both the black hole mass and Newton's constant are constant, hence
the Hawking evaporation process is identical to that in ordinary Einstein
gravity. All that is needed for cosmology then is to transform
the results back into the Jordan-Fierz time frame in which the 
matter is typically understood to evolve. 

If alternatively the evaporation is analyzed in the 
Jordan-Fierz frame,
as in \cite{Barrow,BarrowCarr}, it is necessary to take
into account not only the change of Newton's constant (\ref{Gt})
but also the change of $\Mt$ (\ref{Mt}) arising from the time
dependence of $A(\phi)$. To determine the dependence of the
Hawking luminosity $\Lt$ on $\Gt$ and $\Mt$ we must start with 
the fact that $\Lt$ is determined directly by geometrical 
quantities in the Jordan-Fierz frame\footnote{The surface gravities
in the Jordan-Fierz and Einstein frames are simply related
by $\k = A(\phi_c)\widetilde{\k} $ where $\phi_c$ is the 
asymptotic cosmological value of $\phi$. One way to see this\cite{JK}  is
to note that surface gravity is conformally invariant 
except for the effect due to the different asymptotic normalization 
of the Killing field. The absorption coefficients would in general
not transform in any simple way except for conformally coupled fields, 
however in the adiabatic approximation the conformal factor $A^2(\phi)$
relating the two metrics is constant, so in fact the absorption
coefficients for massless fields are identical in the two frames,
and those for massive fields are related by rescaling the mass.
Thus one can also determine the luminosity $L_*$ directly in the Einstein 
frame. The luminosities $\Lt$ and $L_*$ are related
simply by the transformation of the
energy and time scales, hence we have $L_*=A^2(\phi)\Lt$.} 
: the surface gravity $\widetilde{\k}$ (which determines the 
Hawking temperature 
$\widetilde{T}_H=\hbar\widetilde{\k}/2\pi$) 
and the black hole absorption
coefficients $\widetilde{\Gamma}$. 
The luminosity behaves approximately
as $\Lt\sim \widetilde{\rm Area}\times \widetilde{\k}^4$ which for
a Schwarzschild black hole becomes 
$\Lt\sim {\widetilde{r}_0}^{-2}$
where $\widetilde{r}_0$ is the Schwarzschild radius. Note however that 
$\widetilde{r}_0$ is {\it not} the same as $2\Gt\Mt$.
Being purely geometrical, $\widetilde{r}_0$ must be related to 
the Schwarzschild radius in the Einstein frame $r_{*0}=2G_*M_*$
by the same scale factor that relates the two metrics (\ref{gjf}), 
$\widetilde{r}_0=A(\phi)r_{*0}$.
Using (\ref{Gt}) and (\ref{Mt}) on the other hand we find
$\Gt\Mt=(1 +\a^2)AG_*M_*$, hence evidently 
$\widetilde{r}_0=2\Gt\Mt/(1+\a^2)$.

If $A(\phi)$ decreases by a large amount between the epoch of primordial
black hole formation and today 
then it is possible that the Jordan-Fierz frame 
mass of primordial black holes would 
have been significantly magnified. Such large changes of $A(\phi)$ 
are not inconsistent with observations. According to 
Damour and Pichon\cite{DamourPichon} nucleosynthesis bounds
are consistent with $A_{10\, \rm MeV}/A_{\rm today}$ as large as
150 or even greater.  
Going back further to the early radiation era $A(\phi)$
could in principle have been tremendously larger. Indeed,
Damour and Polyakov argued in a dilatonic model\cite{DamourPolyakov} that, 
as a result of the effect accumulated each
time the temperature drops through the annihilation energy
of a particle species during the radiation era, 
$A(\phi)$ is decreased by a net factor given by
$A_{\rm out}/A_{\rm in}\sim (A_{\rm out}/A_{\rm today})^{-1/F^2}$
where $F = -1.87\times 10^{-4} \kappa^{-9/4}$
and $\kappa$ is a parameter which is naturally of order unity.
Thus $A(\phi)$ could have decreased since the beginning
of the radiation era by a factor of order
$10^{10^8}$ for example if $A_{\rm out}/A_{\rm today} = 100$.
Although the quadratic model 
$\ln A(\phi) \propto(\phi-\phi_{\rm min})^2$ underlying
these calculations should 
not be taken seriously over such a range, the numbers serve to illustrate 
the point that $A(\phi)$ could have been extremely large at the beginning
of the radiation era compared to now.

It is thus conceivable that 
the mass of a small primordial black hole could have been magnified 
by a very large factor. The mass today depends of course not
only on the mass magnification factor but on the initial mass.
Since the Einstein frame mass is unchanged and today the two frames
coincide, this ``mass magnification" would only deserve the name
if the expected masses for primordial black holes were 
fixed in the Jordan-Fierz frame rather than the Einstein frame.
Whether this is true would depend on the physics producing these
black holes. 

The potential mass spectrum of 
primordial black holes will not be discussed here, 
except to point out one fact.  
Carr and Hawking's original estimate\cite{CarrHawking}, that the mass
of primordial black holes formed from collapse of
early universe overdensities might be expected to be of 
order the horizon mass, arose from the agreement
$M_{\rm horizon}\sim M_{\rm Jeans}$ of the upper limit 
horizon mass $M_{\rm horizon}\sim t/G$ and the lower limit 
Jeans mass $M_{\rm Jeans}=(G^3\rho)^{-1/2}$.
This coincidence does not necessarily hold in tensor-scalar
gravity. Since the Jordan-Fierz and Einstein metrics are conformally
related, they define the same particle horizon, so the horizon masses
are related by (\ref{Mt}), 
$\Mt_{\rm horizon}=A^{-1}M_{*{\rm horizon}}\sim A^{-1}(G_*^3\rho_*)^{-1/2}$
(where $\rho_*$ is the Einstein frame
cosmological energy density).  The Jeans mass is naturally 
defined in the Jordan-Fierz frame since the matter is minimally
coupled there. Hence,  
using $\rho_*=A^4\widetilde{\rho}$ and (\ref{Gt}), we have
\begin{eqnarray}
\Mt_{\rm Jeans}&=&(\Gt^3\widetilde{\rho})^{-1/2}\nonumber\\
&=&(1+\alpha^2)^{-3/2}A^{-1}(G_*^3\rho_*)^{-1/2}\nonumber\\
&\sim& (1+\alpha^2)^{-3/2}\Mt_{\rm Horizon}.
\end{eqnarray}
If $\alpha\lsim 1$ then the upper and lower limits still coincide
so the expected mass would be the horizon mass at the time
of formation. If however $\alpha\gg1$, then 
the Jeans mass is much smaller than the horizon mass,
which would allow a much smaller mass for primordial black holes
formed from overdensities than otherwise expected in the standard
scenario\cite{CarrHawking}.  
 
\section*{Acknowledgements}
I would like to thank Doug Armstead for collaboration in the
early stages of this work; John Barrow, Matt Choptuik, Thibault Damour,
Carsten Gundlach, Amos Ori and Patrick Brady for helpful discussions;
and Bernard Carr, Thibault Damour and Carsten Gundlach for useful 
comments on earlier drafts of this paper.
This work was supported in part by the National Science Foundation
under grants No. PHY98-00967 
at the University of Maryland and PHY94-07194 at the Institute for 
Theoretical Physics.


\end{document}